\documentstyle[seceq,epsf]{ptptex}

\title{
The CP-PACS Project and Lattice QCD Results
}

\author{
Y.\ Iwasaki\\
}

\inst{
Center for Computational Physics and Institute of Physics\\
	 University of Tsukuba, Ibaraki 305, Japan
}

\recdate{
}

\abst{
The aim of the CP-PACS project was
to develop a massively parallel computer
for performing numerical research 
in computational physics with primary emphasis
on lattice QCD.
The CP-PACS computer
with a peak speed of 614 GFLOPS with 2048 processors
was completed in September 1996, and has been in full operation since 
October 1996.
We present an overview of the CP-PACS project 
and describe characteristics of
the CP-PACS computer.
The CP-PACS has been mainly used for hadron spectroscopy studies
in lattice QCD.
Main results in lattice QCD simulations are given.
}

\begin{document}

\maketitle

\section{Introduction}
Lattice QCD is a fundamental theory of quarks and gluons which are
constituents of hadrons such as protons and pions.
Numerical studies of lattice QCD have developed significantly
during the past decade in parallel with the development of computers.
Of particular importance in this regard has
been the construction of dedicated QCD computers (see for reviews 
Ref.\citen{review}) and
the move of commercial vendors toward parallel computers in recent years. 
In Japan the first dedicated QCD computer was developed in the QCDPAX
project~\cite{QCDPAX}.
The QCDPAX computer with a peak speed of 14GFLOPS 
is actually the 5th computer in the PAX project~\cite{PAX},
which pioneered the development of parallel computers for scientific and
engineering applications in Japan.

The CP-PACS project was conceived as a successor of the QCDPAX project in 
the early summer of 1991.
The project name CP-PACS is an acronym for Computational
Physics by a Parallel Array Computer System.
The aim of the project was
to develop a massively parallel computer
for carrying out research in computational physics with primary emphasis
on lattice QCD.

In this article after a brief description of lattice QCD and the background
of the project in Sec.2,
we present an overview of the CP-PACS project in Sec.3, 
and describe characteristics of the CP-PACS computer in Sec.4.
The performance of the computer
for lattice QCD applications as well as for the LINPACK benchmark are
also given.
Main results in lattice QCD are given in Sec.5.
Sec.6 is devoted to conclusions.

\section{Lattice QCD and Background of the Project}
Lattice QCD is a fundamental theory of quarks and gluons defined
in terms of the path-integral formalism of quantum theory on a
4-dimensional hyper-cubic lattice.
The lattice spacing plays a role of an ultra-violet cutoff.
The infinite volume limit and
the continuum limit should be taken in order to get physical quantities.

As we have to treat quarks
and gluons relativistically, we have a problem in 4-dimension in
stead of 3-dimension as in solid-state problems.
However, except this difference of dimensionality, it is a statistical
system. Quarks are defined on sites, while gluons on bonds of a
4-dimensional hyper-cubic lattice. Numerical methods we employ are
a Monte Carlo method, a molecular dynamics and a hybrid method of
combination of these methods. 
However, due to this dimensionality, we need a lot of CPU time and 
a large memory size. 

\begin{figure}[th]
\centerline{
\epsfxsize=9.5cm
\epsfbox{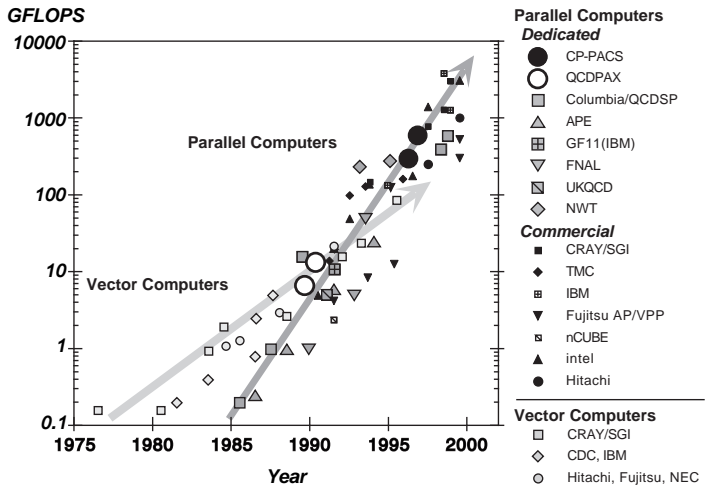}
}
\caption{Recent development of computers in term of theoretical peak speed.}
\label{fig:speed}
\end{figure}


Because of this requirement of high performance computers for numerical
simulations in lattice QCD, dedicated machines have been constructed
in USA, Europe and Japan. 
There are two additional 
reasons why dedicated parallel computers were widely developed for
lattice QCD:
First there is an incentive to perform first-principle calculations
without introducing any approximations based on the fundamental law.
Second there is a spiritual atmosphere in high-energy physics community
to construct a special purposed equipment like an accelerator.
A massively parallel computer is an accelerator for numerical simulations.

Fig.~\ref{fig:speed} shows the recent development of the computers 
in terms of the
theoretical peak speed versus the year when the computer was shipped or
constructed.
Small open symbols are for vector-type supercomputers,
while large and small filled symbols are for dedicated parallel and commercial
parallel computers, respectively.
Open circles with dot are for QCDPAX and filled large circles are for
CP-PACS.
We clearly observe that the rate of the progress for parallel computers is
roughly double that of vector computer and that
a crossover in the peak speed took place from vector to
parallel computers around 1991.
For this development 
dedicated machines for lattice QCD made important roles.

\section{CP-PACS Project}
The CP-PACS Project~\cite{cppacs} aimed at developing a massively parallel 
computer designed
to achieve high performance for numerical research of the major problems of
computational physics,
and it further aimed at significant
progress in the solution of these problems through the application of the 
parallel computer upon completion of its development.

The Project
formally started in April of 1992, and continued for five years, until
March of 1997.
The Project received about 2.2 billion Yen spread 
over the five year period.
%
The Center for Computational Physics was founded in April 1992 at University
of Tsukuba to carry out the Project, as well as to promote
research in computational physics and parallel computer science. The Center
is an inter-university facility open to researchers in academic institutions
in Japan.

The Project members consist of 15 computer
scientists and 18 physicists, as listed in Table~\ref{tab:member}.
As Table~\ref{tab:member} clearly shows, the
CP-PACS Project is a multi-disciplinary effort 
toward the advancement of computational physics
encompassing not only several branches of physics but also
computer science to develop parallel computers best suited for such 
applications.
The Projected was headed by Y. Iwasaki. The development of the CP-PACS
computer was led by K. Nakazawa.

\begin{table}[t]
\setlength{\tabcolsep}{0.25pc}
\newlength{\digitwidth} \settowidth{\digitwidth}{\rm 0}
\catcode`?=\active \def?{\kern\digitwidth}
\caption{CP-PACS Project members}
\label{tab:member}
\begin{center}
\begin{tabular}{lllll}
\hline
{\small hardware}        &{\small software}     &{\small particle}     \
 &{\small astrophysics}  &{\small condensed}\\
 & & {\small physics}    &  &{\small matter} \\
\hline
{\small K. Nakazawa${}^a$}      &{\small I. Nakata${}^e$}          &{\small Y. \
Iwasaki${}^{c}$}   &{\small S. Miyama${}^o$}  &{\small S. Miyashita${}^p$}\\
{\small H. Nakamura${}^b$}	&{\small Y. Yamashita${}^e$}	&{\small A. Ukawa${}^k$}	  &{\small T. Nakamura${}^l$}	&{\small M. Imada${}^q$}\\ 
{\small T. Boku${}^c$}	    &{\small Y. Oyanagi${}^f$}	  &{\small K. Kanaya${}^c$}	 &{\small M. Umemura${}^c$}
&{\small K. Nemoto${}^r$}\\
{\small T. Hoshino${}^d$}	 &{\small T. Kawai${}^g$}	    &{\small S. Aoki${}^k$}	   &{\small
Y. Nakamoto${}^c$}  & {\small A. Oshiyama${}^k$}\\ 
{\small T. Shirakawa${}^d$}&{\small M. Mori${}^h$}	     &{\small T. Yoshie${}^c$} &   & {\small S. Gunji${}^c$}\\
{\small K. Wada${}^e$}	    &{\small Y. Watase${}^i$}	   &{\small M. Okawa${}^l$}\\
{\small M. Yasunaga${}^e$}  &{\small S. Ichii${}^j$}     &{\small N. Ishizuka${}^c$}\\
{\small S. Sakai${}^c$}  &        &{\small M. Fukugita${}^m$}\\
                  &                   &{\small H. Kawai${}^n$}\\
\hline
\multicolumn{5}{l}{\small ${}^a$ Department of Computer Science, University of Electro-Communications}\\[-2pt] 
\multicolumn{5}{l}{\small ${}^b$ Center for Advanced Science and Techology, University of Tokyo}\\[-2pt] 
\multicolumn{5}{l}{\small ${}^c$ Center for Computational Physics, University of Tsukuba}\\[-2pt] 
\multicolumn{5}{l}{\small ${}^d$ Institute of Engineering Mechanics, University of
Tsukuba}\\[-2pt] 
\multicolumn{5}{l}{\small ${}^e$ Institute of Information Sciences and Electronics,
University of Tsukuba}\\[-2pt] 
\multicolumn{5}{l}{\small ${}^f$ Department of Information Science, University of
Tokyo}\\[-2pt] 
\multicolumn{5}{l}{\small ${}^g$ Department of Physics, Keio University}\\[-2pt] 
\multicolumn{5}{l}{\small ${}^h$ Department of Engineering, University of
Tokyo}\\ [-2pt]
\multicolumn{5}{l}{\small ${}^i$ Data Handling Division, KEK}\\[-2pt] 
\multicolumn{5}{l}{\small ${}^j$ Computer Center, University of Tokyo}\\[-2pt] 
\multicolumn{5}{l}{\small ${}^k$ Institute of Physics, University of Tsukuba}\\[-2pt] 
\multicolumn{5}{l}{\small ${}^l$ Numerical Theory Division, KEK}\\ [-2pt]
\multicolumn{5}{l}{\small ${}^m$ Yukawa Institute for Theoretical Physics,
Kyoto University}\\[-2pt] 
\multicolumn{5}{l}{\small ${}^n$ Theory Division, KEK}\\ [-2pt]
\multicolumn{5}{l}{\small ${}^o$ National Astronomial Observatory}\\[-2pt] 
\multicolumn{5}{l}{\small ${}^p$ Department of Physics, Osaka University}\\[-2pt] 
\multicolumn{5}{l}{\small ${}^q$ Institute of Solid State Physics, University of Tokyo}\\[-2pt] 
\multicolumn{5}{l}{\small ${}^r$ Department of Physics, Hokkaido University}\\[-2pt]  
\end{tabular}
\end{center}
\end{table}

A unique feature of the Project is its emphasis on cross-disciplinary
research involving both physicists and computer scientists. This is a 
tradition carried over from the QCDPAX Project~\cite{QCDPAX},
which is the predecessor and  stepping stone for the CP-PACS Project. 
A close collaboration of researchers from
the two disciplines has been both important and fruitful in reaching a
design for the CP-PACS computer which best balances the computational needs 
of physics applications with the latest of computer technologies.

Development of a massively parallel computer requires advanced semiconductor 
technology.  
We selected Hitachi Ltd. as the industrial parter through a formal
bidding process in the early summer of 
1992, and we
worked in a close collaboration for the hardware and software
development of the CP-PACS computer.
The first stage of the CP-PACS computer consisting of 1024 processing
units with a peak speed of 307 GFLOPS was completed in March 1996. An upgrade
to a 2048 system with a peak speed of 614GFLOPS
was completed at the end of September 1996

\section{CP-PACS Computer}

\subsection{Architecture}

The CP-PACS computer is an 
MIMD (Multiple Instruction-streams Multiple Data-streams)
parallel computer with a theoretical peak speed of 614GFLOPS and a
distributed memory of 128 Gbytes. The system consists of 2048
processing units (PU's) for parallel floating point processing and 128
I/O units (IOU's) for distributed input/output processing. These units
are connected in an 8$\times$17$\times$16 
three-dimensional array by a three-dimensional crossbar network. 
The specification of the CP-PACS computer is summarized in
Table~\ref{tab:specification}.
A well-balanced performance of CPU, network and I/O devices
supports the high capability of CP-PACS for massively parallel processing.

\begin{table}[tbh]
\caption{Specification of the CP-PACS computer}
\label{tab:specification}
\begin{center}
\begin{tabular}{ll}
\hline
peak speed &614Gflops(64 bit data)\\
main memory & 128GB\\
parallel architecture&MIMD with \\
                     &distributed memory\\
number of nodes&2048\\
node processor &HP PA-RISC1.1+PVP-SW\\
\hspace*{2mm}\#FP registers&128\\
\hspace*{2mm}clock cycle & 150MHz\\
\hspace*{2mm}1st level cache&16KB(I)+16KB(D)\\
\hspace*{2mm}2nd level cache&512KB(I)+512KB(D)\\
network&3-d crossbar\\
\hspace*{2mm}node array &$8\times 17\times 16^*$\\
\hspace*{2mm}through-put&$$300MB/sec\\
\hspace*{2mm}latency&$2.5 \sim 3.1\,\mu$sec\\
distributed disks&3.5" RAID-5 disk\\
\hspace*{2mm}total capacity&595GB\\
software&\\
\hspace*{2mm}OS&UNIX, micro kernel\\
\hspace*{2mm}language&FORTRAN, C, assembler\\
Size &7.0m(width) $\times$ 4.2m(depth) $\times$ 2.0m(hight)\\
Power dissipation&275 KW maximum\\
\hline
\multicolumn{2}{r}{${}^*$including nodes for disk I/O}
\end{tabular}
\end{center}
\end{table}

The basic strategy we adopted for the design is the usage of a fast RISC
micro-processor for high arithmetic performance at each node and a 
linking of nodes with a flexible network so as to be able to handle a wide 
variety of problems in computational physics.  The
unique features of the CP-PACS computer reflecting these goals are 
represented by the special node processor architecture 
called {\it pseudo vector processor based on slide-windowed
registers} ({\it PVP-SW})\cite{slidewindow} and the choice of a 
three-dimensional crossbar network.  

\subsection{Node processor}

Each PU of the CP-PACS has a custom-made superscalar RISC processor with
an architecture based on PA-RISC 1.1. In large scale computations in 
scientific and engineering applications on a RISC processor, the performance 
degradation occurring when the data size exceeds the cache memory capacity is 
a serious problem.
For the processor of CP-PACS, an enhancement of the architecture called the 
PVP-SW\cite{slidewindow}
was developed to
resolve this problem, while still maintaining upward compatibility with
the PA-RISC architecture.


\subsection{Network}
The 2048 processors are arranged in a three-dimensional $8\times 16\times 16$
array.  
The Hyper Crossbar network is made of crossbar
switches in the $x, y$ and $z$ directions, connected together by an
Exchanger at each of the three-dimensional crossing points of the
crossbar array.
Each exchanger is connected to a PU or IOU.  
Thus
any pattern of data transfer can be performed with the use of at most
three crossbar switches.
Since the network has a huge switching capacity
due to the large number of crossbar switches, the sustained data transfer
throughput in general applications is very high.


\subsection{Performance}
The most CPU consuming part of lattice QCD calculations is
the inversion of a linear equation.
We developed a hand-optimized assembler code for the core
part of the solver.
The performance of the calculation part is 186 MFLOPS per node,
which is 62\% of the peak speed. The percentage of the communication
in the total is 23 \%, which makes the sustained speed for the solver
148 MFLOPS. This is about a half of the theoretical peak speed.

We also measured the performance of the LINPACK benchmark.
The sustained speed for the case of 2048 PU's is 368.2 GFLOPS, 
which is 59.9\% of the theoretical peak speed.
This performance was ranked as number one of TOP 500 
Supercomputers announced in November 
1996.

\section{Physics Results}
\subsection{Hadron Spectrum in Quenched QCD}
Deriving the hadron spectrum from lattice QCD is a milestone
to verify
that QCD is the fundamental theory of quarks and gluons.
Therefore, much effort has been paid to calculate the hadron 
spectrum\cite{spectrum-review}
since 1981 when the first attempt of the hadron spectrum calculation 
was made\cite{first-spectrum}.

A simulation of QCD without approximation requires an enormous computer
time. Therefore, as the first step, the quenched approximation, in which
pair creations and annihilations of quarks in the vacuum are ignored, has
been employed in major simulations of QCD. However, even in the quenched
approximation, it is not easy to obtain precise values of the hadron
spectrum. We have to first control and then estimate various systematic errors 
characteristic of lattice QCD, i.e., the errors due to the infinite volume 
limit and the continuum limit. Moreover, it is technically difficult to
simulate directly at the realistic values of light {\it u} and 
{\it d} quark masses, as the CPU time is proportional to the inverse of the
quark mass. Therefore we have to extrapolate results obtained at relatively
heavy quark masses to the light quark mass. This introduces 
another source of systematic errors. 

In early works, it was difficult
to employ large enough lattices with small enough lattice spacings, mainly
due to limitation of computer power. In particular, all simulations before
1988 employed lattices much smaller than 2 fm which is the size of typical
hadrons. Therefore old calculations suffer from large systematic errors.
Simulations at light enough quark mass were also difficult due to 
algorithm adopted and the speed of computers at that time.

The best calculation prior to the CP-PACS was performed by the GF11
collaboration\cite{GF11} 
in 1992-1993 using their dedicated computer GF11. Performing
systematic extrapolations in terms of quark masses and lattice spacing,
supplemented by corrections from the finite lattice size, they determined
the quenched hadron spectrum in the continuum limit. They concluded that
the hadron masses in  the quenched QCD are consistent with experiment within
their errors, which is typically about 10\%.

As the first physics project on the CP-PACS, we aimed to obtain final
results for the hadron spectrum in the quenched QCD with errors of 
a few \% level and thereby clarify the long standing issue of the 
magnitude of quenching errors. Simulation parameters were chosen
by taking this goal into consideration.

From these simulations together with detailed systematic analyses, we
succeed to determine the quenched hadron spectrum with errors about 1-2 \%
for mesons and 2-3 \% for baryons\cite{quench-letter}.
We were also able to much reduce various systematic errors and estimate 
them. This is crucial to obtain reliable numerical results.
Thus we are able to establish the hadron spectrum in the quenched
QCD. 

\begin{figure}[t]
\begin{center}
\leavevmode
\epsfxsize=8.0cm
\epsfbox{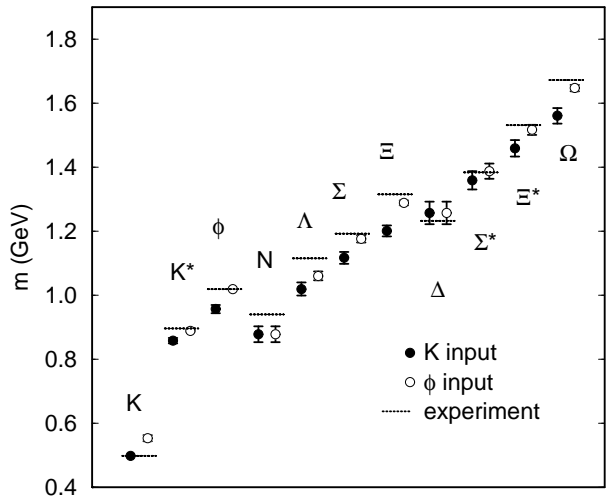}
\vspace{-0.2 cm}
\caption{Hadron masses in quenched QCD, compared with experiment.}
\label{fig:spectrum}
\vspace{-0.2 cm}
\end{center}
\end{figure}

In Fig.\ref{fig:spectrum}, our results for the quenched
 spectrum together with experiment are shown.
The experimental values of the $\pi, \rho$ and $K$ or $\phi$ masses 
are employed to
fix the physical scale and the light quark masses.

Our results unambiguously establish a discrepancy between the quenched
hadron masses and the experimental values, with up to 7$\sigma$ for
several particles. On the other hand, the magnitude of the discrepancy
is at most 10\%, which is consistent with phenomenological estimates of the
quenching error.

\subsection{Hadron Spectrum in Full QCD}
Since the quenched hadron mass spectrum exhibits deviation from experiment,
the next step is to perform calculation of QCD without the quenched 
approximation (the full QCD calculation). As a step toward this goal,
we have started QCD simulations taking into account of effects of
pair creation and annihilation of light {\it u, d} quarks.
We treat the heavier {\it s} quark in the quenched approximation.

Simulations in full QCD need computer power at least 100 times larger
than that in the quenched QCD. Therefore, it is impossible to simply repeat
the simulation in full QCD like that described above in the quenched QCD.
In order to overcome this problem, we adopt an improved action, which
is a lattice action modified in such a way that systematic errors due to
finite lattice spacing is reduced.

We first made a pilot study to investigate the effects of improving
using various improved actions and found that the combination of the 
renormalization-group improved action\cite{RG} 
for gluons and the clover action\cite{clover}
significantly reduces errors due to the finite lattice discretization
over the standard action\cite{comparison}. 
We adopt this combination of improved actions in our production runs.

\begin{figure}[tbh]
\begin{center}
\leavevmode
\epsfxsize=8.0cm
\epsfbox{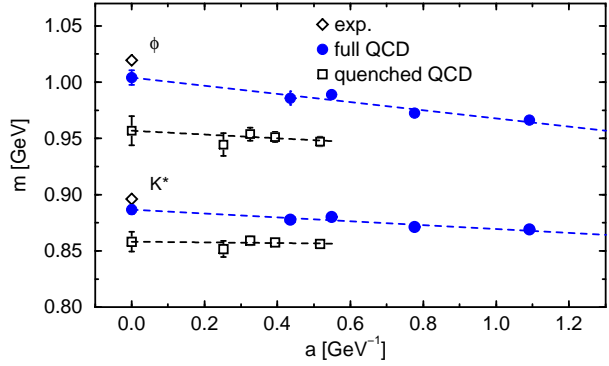}
\vspace{-0.2 cm}
\caption{Meson masses in full QCD compared with those in quenched QCD;
their lattice spacing dependence and continuum limits.}
\label{fig:meson}
\vspace{-0.2 cm}
\end{center}
\end{figure}

A systematic study of the mass spectrum in full QCD is in progress.
We have already found several interesting effects of dynamical quarks in
the hadron spectrum. In Fig.\ref{fig:meson} we compare meson
masses in full QCD with those in the quenched QCD. It clearly
shows that in the continuum limit (the point where the lattice spacing
{\it a} is zero) the discrepancies of $K^*$ and $\phi$ meson masses 
from experiment observed in the quenched QCD are significantly reduced
in full QCD\cite{lat99-kaneko}.

\subsection{Quark Masses}
The masses of quarks are the very fundamental parameters in nature
like the mass of the electron. However, because quarks are confined in
hadrons, one cannot determine their masses directly from experiment.
Usually, their values have been theoretically inferred from experimental
hadron masses using phenomenological models of QCD. 
Lattice QCD is the only known way to determine the masses of quarks
from first principles. 

\begin{figure}[htb]
\begin{center}
\leavevmode
\epsfxsize=7.0cm
\epsfbox{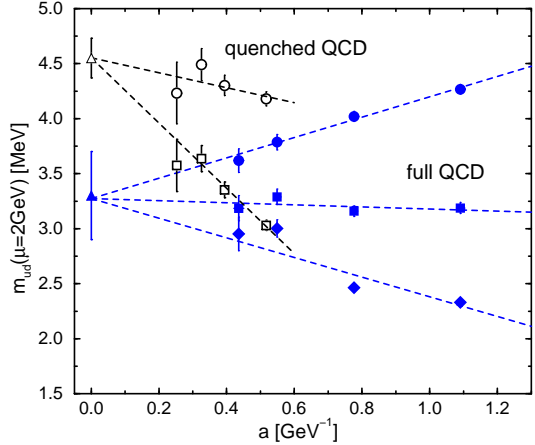}
\vspace{-0.2 cm}
\caption{Mass of the average of {\it u} and {\it d} quarks;
lattice spacing dependence and continuum limit.}
\label{fig:ud-mass}
\vspace{-0.2 cm}
\end{center}
\end{figure}

\begin{figure}[thb]
\begin{center}
\leavevmode
\epsfxsize=8.0cm
\epsfbox{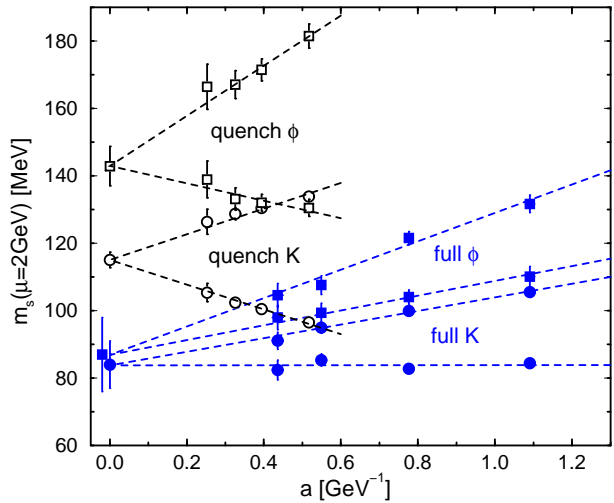}
\vspace{-0.2 cm}
\caption{Mass of the {\it s} quark;
lattice spacing dependence and continuum limit.}
\label{fig:s-mass}
\vspace{-0.2 cm}
\end{center}
\end{figure}

We made systematic calculations of quark masses both in the 
quenched QCD and in full QCD\cite{lat99-kaneko}. 
In Figs.~\ref{fig:ud-mass}
and ~\ref{fig:s-mass}
we show the lattice spacing dependence of the average {\it u, d} quark 
mass and the {\it s} quark mass, respectively. 

On the lattice there are alternative definitions of the quark mass.
Although the values of the quark mass differ depending on the definition
at finite lattice spacing, they extrapolate to a common value in the
continuum limit. The verification of the unique value in the continuum
was first made in the quenched QCD in Ref.\citen{quench-first}.
This verification is important 
because the quark mass should be the fundamental
parameter in QCD.

The {\it s} quark mass is determined using experimental values of either 
$K$ meson mass or $\phi$ meson mass. The {\it s} quark mass in the quenched
approximation depends on the choice of input. This reflects a systematic
error of quenching.

The discrepancy is found to be much reduced in our full QCD calculations.
The values of the {\it s} quark mass from $K$ meson mass or $\phi$ meson mass
are consistent within one standard deviation; 90(10) MeV. This value
is significantly  smaller than that in the quenched QCD; 120-140 MeV.
The value 90(10) MeV for the {\it s} quark mass has a significant
implication for the analysis of the CP violation.

For the clarification of the CP violation in nature we need a theory
like the Kobayashi-Maskawa theory, an experiment result like that
from a B factory,
and also numerical results from lattice QCD. This is a typical example of
cases where results from three fields of
theoretical physics, experimental physics and computational physics
are necessary to solve a problem.

\section{Conclusions}
It was successful to develop a massively parallel computer CP-PACS 
with a peak speed of 614 GFLOPS due to a close collaboration among
physicists, computer scientists and a vendor. The performance of
the computer for physics application is as high as
50 \% of the peak speed
in the case of the core part of lattice QCD programs. 

We are able to obtain intersting and important results in lattice QCD
using the CP-PACS computer: 
1) The hadron spectrum in the quenched QCD has been established.
Our results unambiguously clarify a discrepancy between the quenched
hadron masses and the experimental values, with up to 7$\sigma$ for
several particles. On the other hand, the magnitude of the discrepancy
is at most 10\%, which is consistent with phenomenological estimates of the
quenching error.
2) The discrepancies 
of meson masses from experiment observed in the quenched QCD are 
significantly reduced in full QCD.
3)We have systematically calculate the masses of light quarks in the quenched
QCD and in full QCD. In particular,
the mass of the {\it s} quark in full QCD is 90(10) MeV, which is
much smaller than that previously estimated phenomenologically.

\section*{ACKNOWLEDGEMENTS}

I would like to thank the members of the CP-PACS Project
and the members of the CP-PACS Collaboration for lattice QCD
simulations,
in particular, K. Nakazawa and A. Ukawa
for valuable discussions.
This work is supported in part by
the Grand-in-Aid of the  Ministry of Education, Science and 
Culture (Nos.~08NP0101 and 09304029).

\end{document}